\documentclass[prb,twocolumn,superscriptaddress,amsmath,amssymb,showpacs]{revtex4}

\usepackage{graphicx}
\usepackage{latexsym}
\usepackage{amsmath}
\usepackage{amssymb}
\usepackage{amsfonts}
\usepackage{color}

\newcommand{\ds}{\displaystyle}

\begin{document}
\bibliographystyle{apsrev}

\title{Penetration of the magnetic field into the twinning plane in the type I and II superconductors}

\author{S.V. Mironov}
\affiliation{Institute for Physics of Microstructures, Russian
Academy of Sciences, 603950 Nizhny Novgorod, GSP-105, Russia}
\author{A. Buzdin}
\affiliation{Institut Universitaire de France and University
Bordeaux, LOMA UMR-CNRS 5798, F-33405 Talence Cedex, France}

\date{\today}
\begin{abstract}

It is demonstrated that in the type I and II superconductors with
low-transparent twinning planes (TP) the penetration of external
parallel magnetic field into the region of the twinning plane can
be energetically favorable. In the type I superconductors the
twinning planes become similar to Josephson junctions and the
magnetic field penetrates into the center of the TP in the form of
soft Josephson-like vortices. This leads to increase in the
critical magnetic field values. The corresponding phase diagram in
the parameter plane ``temperature - magnetic field'' essentially
differs from the one obtained without taking the finite value of
the magnetic field near the TP into account. Comparison
between obtained phase diagrams and experimental data for
different type I superconductors can allow to estimate the value
of the TP transparency, which is the only fitting parameter in our
theory.

\end{abstract}


\maketitle

The phenomenon of the twinning plane superconductivity (TPS) has
been the subject of intensive investigations during last three
decades (see Ref.~\onlinecite{Khlyustikov} for review). A twinning
plane (TP) may produce more favorable conditions for the
superconducting nucleation compared with a bulk crystal and a
superconducting layer localized on the TP can appear even above
the bulk critical temperature $T_c$. Recently the interest to the
physics of twins in superconductors was renewed since it was shown
experimentally that TP affect the properties of many relatively
new superconductors which belong to the pnictide
family\cite{Ni,Tanatar,Chu,Kalisky,Kirtley,Prozorov}. In these
superconductors twinning planes can enhance locally the superfluid
density\cite{Kalisky,Kirtley} or influence the vortex
pinning\cite{Prozorov, KaliskyVortices}. In particular it was
demonstrated\cite{KaliskyVortices} that in ${\rm
Ba(Fe_{1-x}Co_x)_2As_2}$ twinning planes repulse vortices and act
as strong barriers for vortex motion. Thereupon the theoretical
investigations of the magnetic properties of the TP in
superconductors are of current importance.

Twinning planes can effectively screen parallel magnetic field
which leads to the increase in the value of the critical field as
a function of temperature. The corresponding phase diagrams
$H_c(T)$ for the absolutely transparent TP were studied
theoretically within the phenomenological Ginzburg-Landau
formalism both for the type I \cite{Averin, Khvorikov, Mishonov,
Osborn, Indekeu} and the type II superconductors\cite{Khvorikov}.
In particular for the ultra type I superconductors it was shown
\cite{Mishonov, Osborn, Indekeu} that for small but finite values
of the Ginzburg-Landau parameter $\kappa$ the magnetic field
penetration into the superconducting area leads to the corrections
to the TPS free energy proportional to $\kappa^{(n+1)/2}$
($n=0,1,...$) and in practice only the term $\propto\kappa^{1/2}$
plays an important role while the terms of the order
$\kappa^{3/2}$ and higher can be neglected since they weakly
contribute to the resulting $H_c(T)$ diagram. The resulting dependencies $H_c(T)$ do not contain any fitting
parameters, which allowed their quantitative experimental
verification for concrete superconductors \cite{Khlyustikov_JETP,
Moskvin}. In some cases\cite{Kozhevnikov_1,Kozhevnikov_2} the local enhancement of superconductivity may occur near the sample surface. The upper critical field for this situation was considered in Ref.~\onlinecite{Kuptsov}. However for the type I-Sn the superconductivity in magnetic field should appear as I order transition (except very narrow region near the critical temperature) and this field corresponds to the overcooling of the normal phase.

Practically for TP with finite electron transparency the standard Ginzburg-Landau free energy functional should be generalized by inserting additional term which breaks the requirement of the order parameter continuity at TP.\cite{Andreev, Geshkenbein} The influence of the finite TP transparency on the upper critical magnetic field in the type II superconductors was analyzed in Ref.~\onlinecite{Samokhin}. At the same time for superconductors of the I type it was found that in case of low transparent TP the essentially asymmetric distributions of the order parameter relative to the TP can become energetically more favorable than symmetric ones.\cite{Indekeu_Theory} The striking prediction of this paper is that under certain conditions the order parameter is nonzero only at one side of the TP while at another side it should be zero.

We would like to point out that all theoretical results which have been obtained up to now
are based on the assumption that superconducting nucleus localized near
TP is screening the magnetic field effectively: it was believed
that in the type I superconductors the magnetic field can
penetrate only into the region which is far from the TP (the
magnetic field value at the TP is exponentially small) while in
the type II superconductors the magnetic field value has its
minimum at the TP.

In the present paper we show that for both type I and II
superconductors with low-transparent TP the magnetic field can
fully penetrate into the center of the twinning plane and the
corresponding state is energetically favorable. For the type II
superconductors this fact results in small corrections to the
magnetic susceptibility only. At the same time for the type I
superconductors the magnetic field penetration into the center of
the TP leads also to essential changes in the dependence of the
critical magnetic field on temperature due to the negative
contribution to the free energy, which has the order of $\kappa$. Note that the obtained solutions have lower energy than the ones found in Ref.~\onlinecite{Indekeu_Theory}. At the same time the corresponding profiles of the order parameter are symmetric relative to the TP.

Let us consider a bulk superconducting sample with a single
twinning plane at $z=0$. The external magnetic field ${\bf H}$ is
assumed to have only the $y$ component. We choose the
corresponding vector potential in the form $A_x(z)=Hz$ so that the
order parameter $\psi$ depend only on $x$ and $z$. We will use the
standard Ginzburg-Landau free energy functional to describe the
local enhancement of the superconductivity on the
TP\cite{Mishonov, Samokhin}
\begin{widetext}
\begin{equation}\label{Free_Energy}
G=\int
dxdz\left\{\frac{\hbar^2}{4m}\left|\left(\nabla-\frac{2ie}{\hbar
c}{\bf
A}\right)\psi\right|^2+a\left|\psi\right|^2+\frac{b}{2}\left|\psi\right|^4+\frac{\left({\bf
B}-{\bf
H}\right)^2}{8\pi}+\frac{\hbar^2}{4m}\left[\frac{8}{\rho}\left|\psi_{+}-\psi_{-}\right|^2-\frac{1}{2\xi_s}
\left(\left|\psi_{+}\right|^2+\left|\psi_{-}\right|^2\right)
\right]\delta(z)\right\},
\end{equation}
\end{widetext}
where $a=\alpha(T-T_{c})$, ${\bf B}={\rm rot}{\bf A}$, $\rho$ is
a phenomenological constant describing the finite transparency of
the TP, $\psi_{\pm}=\psi(x,y,\pm 0)$ and the value $\xi_s$ will be
defined below. Let us also introduce the temperature dependent
coherence length as $\xi(T)=\hbar/\sqrt{4m\alpha(T-T_{c})}$.

For the further analysis it is convenient to rewrite the functional
(\ref{Free_Energy}) in dimensionless variables
\begin{equation}\label{Dim_Variables}
\begin{array}{c}
{\ds
t=\frac{T-T_c}{T_s-T_c},~H_s=H_c(t=-1),~\xi_s=\xi(t=-1),}\\{\ds
\psi_s=\psi_0(t=-1),~\tilde{\psi}=\frac{\psi}{\psi_s},~\tilde{x}=\frac{x}{\xi_s},~\tilde{z}=\frac{z}{\xi_s},
}\\{\ds h=\frac{H}{H_s},~\tilde{\bf A}=\frac{\bf A}{\kappa H_s
\xi_s},~r=\frac{\rho}{\xi_s},~G_s=\frac{\xi_s H_s^2}{8\pi}}.
\end{array}
\end{equation}
Here $T_s$ is the critical temperature of the superconductor with
the twinning plane ($T_s>T_c$), $\psi_0(T)=\sqrt{|a|/b}$ and
$H_c(T)=\sqrt{4\pi a^2/b}$ are the values of the order parameter
wave function and the critical filed for the bulk superconductor
without twinning plane; $\kappa=mc\sqrt{b}/\sqrt{2\pi}e\hbar$ is
the Ginzburg-Landau parameter. In what follows we will omit the
tildes since we will consider only dimensionless
expressions. Then the functional for the free-energy per unit
length along the TP reads as
\begin{equation}\label{Free_Energy_Dimless}
\begin{array}{c}{\ds
G=G_s\int dxdz\left\{2\left|\frac{\partial\psi}{\partial
z}\right|^2+2\left|\left(\frac{\partial}{\partial x}-\frac{iA
}{\sqrt{2}}\right)\psi\right|^2+2t\left|\psi\right|^2\right.}\\{\ds
\left.+\left|\psi\right|^4+\left[\frac{2}{r}\left|\psi_{+}-\psi_{-}\right|^2- 2\left(\left|\psi_{+}\right|^2+\left|\psi_{-}\right|^2\right)
\right]\delta(z)\right.}\\{\left.+\left(\kappa\frac{dA}{dz}-
h\right)^2\right\}}.
\end{array}
\end{equation}

Let us start with the case of the absolutely opaque TP
($r=\infty$) in the ultra type I superconductor ($\kappa\ll 1$)
and consider the order parameter wave function in the form
$\psi(x,z)=\varphi(z){\rm exp}\{i\theta(x)\}$. Varying the
functional (\ref{Free_Energy_Dimless}) by $\psi^{*}$ and $A$ for
$z\not=0$ we obtain the equations
\begin{equation}\label{System}
\left\{
\begin{array}{l}
{\ds -\partial_z^2
\varphi+t\varphi+\varphi^3+\varphi\left(A/\sqrt{2}-\partial_x\theta\right)^2=0,}\\{\ds
\partial_z^2 A=(\sqrt{2}\varphi^2/\kappa^2)\left(A/\sqrt{2}-\partial_x\theta\right)}
\end{array}\right.
\end{equation}
with the boundary conditions $\partial_z\varphi_{+}=-\varphi_{+}$
and $\partial_z\varphi_{-}=\varphi_{-}$. Note that the system
(\ref{System}) has the first integral
\begin{equation}\label{First_Int}
(\partial_z\varphi)^2-t\varphi^2- \frac{\varphi^4}{2}-\varphi^2\left(\frac{A}{\sqrt{2}}-
\partial_x\theta\right)^2+\frac{\kappa^2}{2}(\partial_z A)^2=\frac{h^2}{2}.
\end{equation}
The modulus of the order parameter on the TP is given by
\cite{Mishonov} $\varphi_0^2=(1-t)+\sqrt{(1-t)^2-h^2}$. In what
follows we will assume that the magnetic field penetration into
the center of the TP weakly affects the order parameter modulus
near the TP. The validity of this assumption will be discussed
below. Also we would like to mention that the order parameter
decay length $l$ is proportional to $t^{-1/2}$ while the magnetic
field penetration length is proportional to $\kappa/\sqrt{1-t}$ and for
$\kappa\ll 1$ and $(1-t)\gg \kappa^2$ it is much less than $l$. This allows one to
consider the constant order parameter value in the region near the
center of the TP where the local magnetic field
$b=\kappa\partial_zA$ is nonzero. We will assume that
$b(z=\pm\infty)=h$ and $b(z=0)=h_0$, where $h_0\leq h$. Then from
the solution of the equation (\ref{System}) on the vector
potential we obtain
\begin{equation}\label{B_Theta}
b(z)=h_0{\rm
exp}\left\{-\frac{\varphi_0}{\kappa}|z|\right\},~\partial_x\theta=\frac{{\rm
sign} z}{\sqrt{2}}\frac{h_0}{\varphi_0},
\end{equation}
The resulting correction to the free energy value per unit length
along the $x$ axis has the form
$G=G_0+G_s(2\kappa/\varphi_0)\left(h_0^2-2hh_0\right)$, where
$G_0$ is the free energy obtained without taking the penetration
of the magnetic field to the TP into
account\cite{Mishonov,Indekeu}
\[\frac{G_0}{G_s}=\left[4\sqrt{2}\int\limits_0^{\varphi_0}\sqrt{\varphi^4+2t\varphi^2+h^2}d\varphi-4\varphi_0^2-2.06\sqrt{\kappa
h^3}\right].\] From the obtained dependence $G(h_0)$ one can see
that its minimum corresponds to the case $h_0=h$ and the resulting
free-energy value is
\begin{equation}\label{G_min_r_infty}
G=G_0-G_s(2\kappa/\varphi_0)h^2.
\end{equation}
Thus it is energetically favorable for magnetic field to penetrate
fully into the TP center. Then solving numerically the inequality
$G\leq0$ we obtain the temperature dependence of the critical
magnetic field $h_c(t)$, which is shown in Fig.~1 (red solid
curve).

\begin{figure}[t!]
\includegraphics[width=0.5\textwidth]{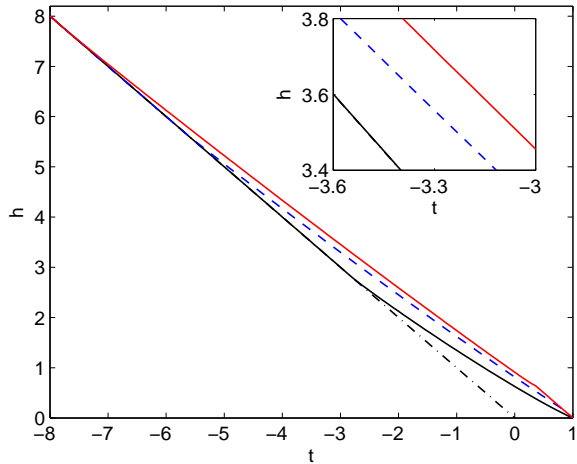}
\caption{(Color online) Phase diagram of the type I
superconductors with absolutely opaque TP in the parallel magnetic
field. Penetration of the magnetic field into the TP leads to
increase in the critical magnetic field $h_c(t)$ (red solid curve)
compared with the one calculated with the assumption of
exponentially small magnetic field on the TP\cite{Mishonov} (blue
dashed curve). In the inset we show the fragment of these
dependencies in more detail. In our calculations we took
$\kappa=0.13$ corresponded to tin. Also critical magnetic fields
of the superconductor with $\kappa=0$ (black solid curve) and for
a bulk superconductor (black dashdot curve) are shown.}
\label{R_0}
\end{figure}

\begin{figure}[t!]
\includegraphics[width=0.5\textwidth]{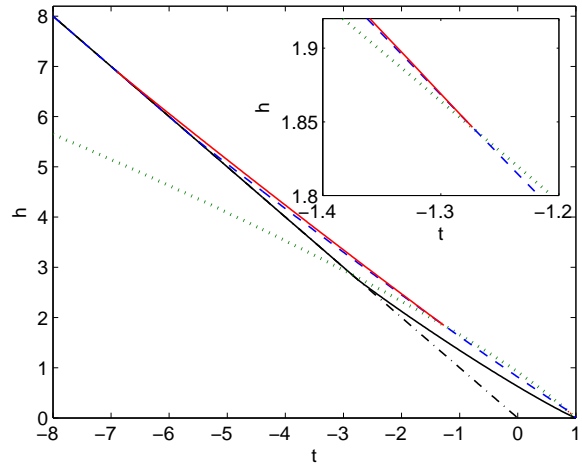}
\caption{(Color online) Phase diagram of the type I
superconductors with low-transparent TP ($r=25$) in the parallel
magnetic field. Penetration of the magnetic field into the TP is
energetically favorable for temperatures below the point where the
dependence of the critical Josephson magnetic field $h_{cJ}(t)$
(green dotted curve) crosses the critical field dependence
calculated with the assumption of the exponentially small magnetic
field on the TP\cite{Mishonov} (blue dashed curve). In this
temperature region the resulting dependence of the critical
magnetic field is shown with red solid curve. In our calculations
we took $\kappa=0.13$ (tin).}
\end{figure}

Note that the magnetic field (\ref{B_Theta}) does not affect the
value of the order parameter at the TP. Indeed substituting the
corresponding vector potential at $z=0$ to the first integral
(\ref{First_Int}) and one can obtain that at the TP the order
parameter
$\varphi^2(z=0)=2(1-t)-h^2/\varphi_0^2\equiv\varphi_0^2$.

The expression for the local magnetic field (\ref{B_Theta}) allows
one to calculate the correction to the magnetic moment of the TP,
which has the form (here we consider only the correction due to
the magnetic field (\ref{B_Theta}) and restore the dimension of
the expression)
\begin{equation}\label{Delta_M}
\Delta M_{(I)}
=\int\limits_{-\infty}^{\infty}\frac{B(z)}{4\pi}dz=\frac{\kappa
H\xi_s}{2\pi\varphi_0}.
\end{equation}
Comparing this value in the limit $h\to 0$ with the diamagnetic
moment $M_{d(I)}$\cite{Khlyustikov} due to the expulsion of the
magnetic field from the superconducting region we obtain that the
ratio $\eta_{(I)}=\left|\Delta
M_{(I)}/M_{d(I)}\right|=\kappa\left({\rm
ln~}\kappa^{-1}\right)^{-1}\sqrt{t/2(1-t)}$. Practically
$\eta_{(I)}\ll 1$ for all temperatures where the TPS has the I
type. Indeed, for example, for tin with $\kappa=0.13$ the
condition $\eta_{(I)}\sim 1$ gives $(1-t)<2\cdot10^{-3}$. For such
small values of $(1-t)$ the obtained results are not applicable
since for $(1-t)<2.7\cdot10^{-2}$ the superconducting phase
transition is of the II type\cite{Khlyustikov}.

Now let us generalize the obtained results for the case of TP with
finite but small transparency (we will assume that $r^{-1}\ll 1$).
For finite $r$ the TP is similar to the Josephson junction with
corresponding Gibbs free energy $G_r/G_s\propto r^{-1}+O(r^{-2})$.
Note that the boundary conditions for the order parameter $\psi$
at $z=0$ should be modified in order to describe the transparent
TP\cite{Samokhin} so that
\begin{equation}\label{Boundary}
\begin{array}{c}{\ds
\partial_z
\psi_{+}=-\psi_{+}+r^{-1}(\psi_{+}-\psi_{-}),}\\{\ds \partial_z
\psi_{-}=\psi_{-}+r^{-1}(\psi_{+}-\psi_{-}).}
\end{array}
\end{equation}
Note that the corrections to the order parameter modulus $|\psi|$
due to the changes in the boundary conditions have the order of
$r^{-1}$. The corresponding corrections to the free energy have at
least the order of $r^{-2}$ and can be neglected. Thus in what
follows we will assume that for low-transparent TP the spatial
distribution of the order parameter module $\varphi(z)$ is the
same as in case of the absolutely opaque TP. Then it is easy to
obtain the expression for the Josephson free energy of the TP,
which has the form $G_r/G_s=(4\varphi_0^2/r)\int\left(1-{\rm
cos}\Delta\theta\right)dx$, where
$\Delta\theta=\theta_{+}-\theta_{-}$ and
$\theta_{\pm}=\theta(x,\pm 0)$.

It is naturally to expect that the magnetic field can penetrate
into the TP center in the form of soft Josephson-like vortices.
Indeed the magnetic field at the TP is defined by the phase
difference on the two sides of the TP and has the form
$h_0(x)=(\varphi_0/\sqrt{2})\partial_x\Delta\theta$ while
dimensionless Josephson current through the TP can be written as
\begin{equation}\label{Josephson_Current}
j_z(x)=\frac{\sqrt{2}\varphi_0^2}{\kappa r}{\rm sin}\Delta\theta
\end{equation}
(here we use the value $cH_s/4\pi\xi_s$ as the unit of current).
Then substituting $h_0(x)$ and $j_z(x)$ into the Maxwell equations
one can obtain the analog of the Ferrell-Prange equation
$\partial_x^2\Delta\theta=\lambda_J^{-2}{\rm sin}\Delta\theta$,
where $\lambda_J=(\kappa r/2\varphi_0)^{1/2}$ is the Josephson
penetration depth. It is convinient to introduce new dimensionless
coordinate $x'=x\sqrt{2\varphi_0/\kappa r}$. Then the magnetic
part of the Helmholtz free energy $F=G+2BH$ per unit length can be
represented in the standard form
\begin{equation}\label{G_r_Helmholtz}
F=G_s\frac{4\varphi_0^2}{r}\int\left\{\left(1-{\rm
cos}\Delta\theta\right)+\frac{1}{2}\left(\partial_{x'}\theta\right)^2\right\}dx'.
\end{equation}
From the expression (\ref{G_r_Helmholtz}) it is easy to obtain the
value of the Josephson critical field $h_{cJ}$ which is the
minimal field of the vortex penetration into the junction
\cite{Kulik}. The expression for $h_{cJ}$ reads as
\begin{equation}\label{Hc1J}
h_{cJ}(t)\approx\frac{4}{\pi}\frac{1}{\sqrt{\kappa
r}}\left[2(1-t)\right]^{3/4}.
\end{equation}

To describe the temperature dependence of the critical magnetic
field which corresponds to field penetration into the TP we use
the results of Ref.~\onlinecite{Kulik}, where the averaged
magnetization $M_{J}$ of the Josephson junction and the
corresponding Helmholtz energy $F_{J}$ are calculated as implicit
functions of the external magnetic field $h$. Substituting these
dependencies into the Gibbs free energy $G$ of the TP and
performing numerical calculations we obtain the dependence of the
critical magnetic field on temperature $h_c(t)$ which is shown in
Fig.~2 (red solid curve). Obviously this dependence exceeds the
dependence $h_{c0}(t)$ corresponded to the condition $G_0=0$ (see
blue dashed curve in Fig.~2) only for temperatures where
$h_{c0}(t)>h_{cJ}(t)$. Note that the obtained dependence
$h_c(t)$ depends only on one fitting parameter $r$. Thus we hope
that high-accuracy measurements of the critical magnetic fields
can allow to estimate the values of the twinning plane
transparencies for different type I superconductors.

Note also that the magnetic field penetration into the center of
the TP can be detected experimentally in the Josephson current
measurements. Let us consider the pair of contacts which are
positioned parallel to the TP at a distance which is much less
than a width of superconducting region $l\propto t^{-1/2}$ near
the TP. Then the Josephson current through these contacts would be
extremely sensitive to the magnetic field in the center of the TP.
Indeed without magnetic field the averaged over the TP length
Josephson current can be nonzero if $\Delta\theta\not= 0$ (see
(\ref{Josephson_Current})). Overwise in case of the magnetic field
penetration the phase difference would increase with the increase
in $x$ and the corresponding averaged Josephson current through
the TP would be negligibly small.

For the type II superconductors the fact of full penetration of
the magnetic field into the center of the TP doesn't lead to any
substantial consequences since the TP in this case weakly screen
the external magnetic field. To calculate the correction to the
magnetic susceptibility one can use the approach from
Ref.~\onlinecite{Khvorikov}. We will restrict ourselves for the
case of ultra type II superconductors with $\kappa\gg 1$,
$r=\infty$ and the temperatures in the range $0<t<1$. In this case
the magnetic field $b(z)$ slightly differs from the external field
$h$ which allows one to consider the field profilein the form
$b(z)=h+\delta b(z)$, where $|\delta b(z)|\ll h$. Then the value
$\delta b(z)$ satisfies the equation $(u^2-1)\partial^2_u \delta
b=2h/\kappa^2$, where $u={\rm coth\left(\sqrt{t}|z|+p/2\right)}$
and $p={\rm ln}[(1+\sqrt{t})/(1-\sqrt{t})]$. The boundary
conditions in the case of full penetration of the magnetic field
into the TP center are $\delta b(u=1)=0$ and $\delta
b(u=t^{-1/2})=0$. Note that under the assumption that the magnetic
field has its minimum on the TP the last condition should be
replaced with\cite{Khvorikov} $\partial_u \delta b(u=t^{-1/2})=0$.
The exact solution of the equation for $\delta b$ allows one to
obtain the correction to the magnetic susceptibility due to the
penetration of the magnetic field into the TP. This correction has
the form (in dimensional units)
\begin{equation}\label{DM_II}
\Delta
M_{(II)}=\frac{H\xi_s}{\kappa^2}\frac{1}{\pi(1-\sqrt{t})}{\rm
ln}^2\left(\frac{1+\sqrt{t}}{2\sqrt{t}}\right).
\end{equation}
Note that at $t\to 0$ the correction $\Delta M_{(II)}\propto {\rm
ln}^2(t^{-1})$ and is negligibly small since it has weak
singularity compared with the full diamagnetic moment $M_{d(II)}$
of the TP, which diverges like $t^{-1/2}{\rm ln}^2(t^{-1})$ (see
Ref.~\onlinecite{Khvorikov}). At $t\to 1$ the correction $\Delta
M_{(II)}\propto (1-t)$ and is also small.

Thus we have shown that in superconductors with twinning planes
with low transparency the penetration of the parallel magnetic
field into the twinning plane is energetically favorable. For the
type I superconductors this leads to the essential increase of the
critical magnetic field and to the broadening of the temperature
range where the TPS can exist. Our theory does not contain
any fitting parameters except $r$, so this can provide an
opportunity to estimate the $r$ values for different type I
superconductors on the basis of the critical magnetic field
measurements.

%

The authors thank A.S. Mel'nikov for many useful discussions and reading the manuscript. This work was supported by the European IRSES program
SIMTEC, French ANR ''SINUS", the RFBR, RAS under the Program
``Quantum physics of condensed matter", the ``Dynasty'' Foundation
and FTP ``Scientific and educational personnel of innovative
Russia in 2009--2013".

\end{document}